\documentclass[10pt,journal,compsoc]{IEEEtran}

\usepackage{cite}
\usepackage{graphicx}
\usepackage{hhline}
\usepackage{comment}
\usepackage[dvipsnames]{xcolor}
\usepackage[colorlinks=false, linkcolor=blue, citecolor=blue, bookmarks=true,pagebackref=false,hidelinks=true]{hyperref} 
\usepackage{hyphenat}
\usepackage{subcaption}
\usepackage{float}
\usepackage{caption}
\usepackage{fancybox}
\usepackage{array,booktabs}
\usepackage{marvosym} 
\usepackage{textcomp} 
\usepackage{tablefootnote} 
\usepackage{balance}
\usepackage{multirow}
\usepackage{soul}
\usepackage[flushleft]{threeparttable} 

\definecolor{mycolor}{RGB}{0, 191, 255} 

\newcommand{\rqbox}[1]{
	\begin{center}
		\vspace{-0.1cm}
		\cornersize{.2}
		\setlength{\fboxsep}{7pt}
		\ovalbox{\begin{minipage}{3.3in}
				{\em #1}
		\end{minipage}}
		\vspace{-0.1cm}
	\end{center}
}

\begin{document}

\title{Does the hiding mechanism for Stack Overflow comments work well? No!}

\author{Haoxiang~Zhang,
	Shaowei~Wang,
	Tse-Hsun~(Peter)~Chen,
	and~Ahmed~E.~Hassan,~\IEEEmembership{Fellow,~IEEE}
	\IEEEcompsocitemizethanks{\IEEEcompsocthanksitem H. Zhang, S. Wang and A. E. Hassan are with the Software Analysis and Intelligence Lab (SAIL), Queen's University, Kingston, Ontario, Canada.\protect\\
		E-mail: {hzhang,shaowei,ahmed}@cs.queensu.ca
		\IEEEcompsocthanksitem T. Chen is with the Department of Computer Science and Software Engineering, Concordia University, Montreal, Quebec, Canada.\protect\\
		E-mail: peterc@encs.concordia.ca
		\IEEEcompsocthanksitem Shaowei Wang is the corresponding author.
	}
}

\IEEEtitleabstractindextext{
	\begin{abstract}
Stack Overflow has accumulated millions of answers to various questions. Many of these answers have associated comments that can assist developers. Informative comments can strengthen their associated answers (e.g., providing additional information). Currently, Stack Overflow applies a mechanism to sort all comments under an answer in a descending order by their scores (i.e., up-votes), then by their creation time. This mechanism hides comments that are ranked beyond the top 5. Stack Overflow aims to display more informative comments (i.e., the ones with higher scores) and hide less informative ones using this mechanism. As a result, 4.4 million comments are hidden under their answer threads, possibly including informative comments although they are ranked below the top 5 comments. Therefore, it is very important to understand how well the current comment hiding mechanism works. In this study, we investigate whether the mechanism can effectively deliver informative comments while hiding uninformative comments. We find that: 1) Hidden comments are as informative as displayed comments; more than half of the comments (both hidden and displayed) are informative (e.g., providing alternative answers, or pointing out flaws in their associated answers). 2) The current comment hiding mechanism tends to rank and hide comments based on their creation time instead of their score in most cases due to the large amount of tie-scored comments (e.g., 87\% of the comments have 0-score). 3) In 97.3\% of answers that have hidden comments, at least one comment is hidden while there is another comment with the same score is displayed (i.e., we refer to such cases as unfairly hidden comments). Among such unfairly hidden comments, the longest unfairly hidden comment is more likely to be informative than the shortest unfairly displayed comments. Our findings suggest that Stack Overflow should consider adjusting their current comment hiding mechanism, e.g., displaying longer unfairly hidden comments to replace shorter unfairly displayed comments. We also recommend that users examine all comments, in case they would miss informative details such as software obsolescence, code error reports, or notices of security vulnerability in hidden comments.
	\end{abstract}
}

\maketitle

\IEEEdisplaynontitleabstractindextext

\IEEEpeerreviewmaketitle

\IEEEraisesectionheading{\section{Introduction}\label{sec:introduction}}
\IEEEPARstart{S}{tack} Overflow is a Q\&A platform that is widely used by software developers to find answers to their programming questions. It has over 50 million visitors in a single day\footnote{Data obtained on Jan 31, 2018, from \url{https://insights.stackoverflow.com/survey/2018/}}. In the archived data provided by Stack Overflow in September 2017, there are 22.7 million answers to questions in various domains related to software development. These answers offer developers with solutions to address their questions. This collection of answers enables developers to learn and share valuable programming knowledge.

An answer creates a starting point for a discussion related to a question. Users can post comments under the answer. We consider that an answer thread is composed of an answer and all the comments under the answer. Comments can provide additional information to support their associated answer, or even point out issues in the answer, such as the obsolescence of answers~\cite{Zhang:2018}.

In order to keep each answer thread compact, Stack Overflow implements a comment hiding mechanism to only display the top 5 comments at most. Aiming at displaying the more informative comments and hiding the less informative ones, the mechanism first ranks the comments based on their scores. When multiple comments have the same score, they are then ranked by their creation time. To read the hidden comments, users need to click a link under the last displayed comments. The more comments are posted under an answer, especially those answers that attract large user traffic
, the larger the proportion of hidden comments for an answer.

We observe that 4.4 million comments are hidden as of September 2017. As once commented by David Fullerton (the president of Stack Overflow), ``\textbf{\textit{if I have to click that link every time just in case there's something useful in the comments, haven't we failed?}}''\footnote{\url{https://meta.stackexchange.com/posts/comments/653443/}}. In other words, it is essential for Stack Overflow to display the most useful comments and hide the less useful ones. In addition, hidden comments are not indexed by Google\footnote{\url{https://meta.stackexchange.com/a/304906/}}, which also hinders the accessibility to the information in comments for answer seekers. Therefore, it is important to understand what users are actually discussing in both hidden and displayed comments, and how well the comment hiding mechanism works. Does this mechanism, in fact, display the more informative comments and hide the less informative comments as expected? By answering these questions, we wish to provide insights to improve the current comment hiding mechanism to make it easier for developers to retrieve information on Stack Overflow.

In this paper, we study 22.7 million answers and all of their 32.2 million associated comments. We first study whether the displayed comments are more informative than the hidden comments. In other words, is the comment hiding mechanism actually hiding less informative comments? We investigate:

\begin{itemize}
\item \textbf{RQ1: What are the characteristics of both hidden and displayed comments?}\\
We find that hidden comments have the same amount of text as displayed comments. Hidden comments have a richer vocabulary and add a greater variety of textual content to their associated answers than that of displayed comments to the same answer.

\item\textbf{RQ2: What do users discuss in hidden and displayed comments?}\\
Based on our qualitative study, we find that more than 70\% of the comments (both hidden and displayed) are informative for the discussion, such as providing alternative answers, or pointing out flaws for answers.
\end{itemize}

From the previous qualitative study, we observe that informative comments can be hidden under the current comment hiding mechanism. To further understand the reason for such cases, we perform an empirical analysis to evaluate the efficacy of the comment hiding mechanism by answering the following two RQs:

\begin{itemize}
\item \textbf{RQ3: How effective is the comment hiding mechanism?}\\
The comment hiding mechanism does not work effectively. Instead of giving priority to highly scored comments, it gives priority to early comments since the comment hiding mechanism fails to consider a very common case: multiple comments may have the same score (i.e., tie-scored comments). More specifically, in 97.3\% of the answers that have hidden comments, at least one comment is hidden (i.e., unfairly hidden comment), while other comments with the same score are displayed (i.e., unfairly displayed comments).


\item \textbf{RQ4: What are the characteristics of unfairly hidden comments?}\\
Based on our qualitative study, we obverse that the longest unfairly hidden comments are more likely to be informative than the shortest unfairly displayed comments (when the shortest unfairly displayed comments are less than 50 characters).
\end{itemize}

Based on above findings, we suggest that Stack Overflow enhances their comment hiding mechanism to better handle tie-scored comments which represent 92.9\% of hidden comments. Instead of simply ranking comments by their score then their creation time, the mechanism needs to introduce a higher priority for more informative comments. 
For example, Stack Overflow can replace the shortest unfairly displayed comments with the longest unfairly hidden comments (i.e., with tied score). In addition, we encourage users to read through all comments (including hidden comments) in case any further correction/improvement is made by such comments, such as observations of answer obsolescence, security vulnerabilities, and errors.

\textbf{Paper Organization:} The rest of the paper is organized as follows. Section~\ref{background} introduces the background of Stack Overflow's comment hiding mechanism. Section~\ref{data} details the dataset used in this study. Section~\ref{study1} and Section~\ref{study2} present the case study results. Section~\ref{discussion} discusses our findings and provides actionable suggestions. Section~\ref{validation} discusses potential threats from our case study. Section~\ref{relatedwork} surveys prior work related to our study. Finally, Section~\ref{conclusion} concludes our study.

\section{Background}
\label{background}
\subsection{Commenting on Stack Overflow}

\begin{figure}[ht]
	\centering\includegraphics[width=\columnwidth]{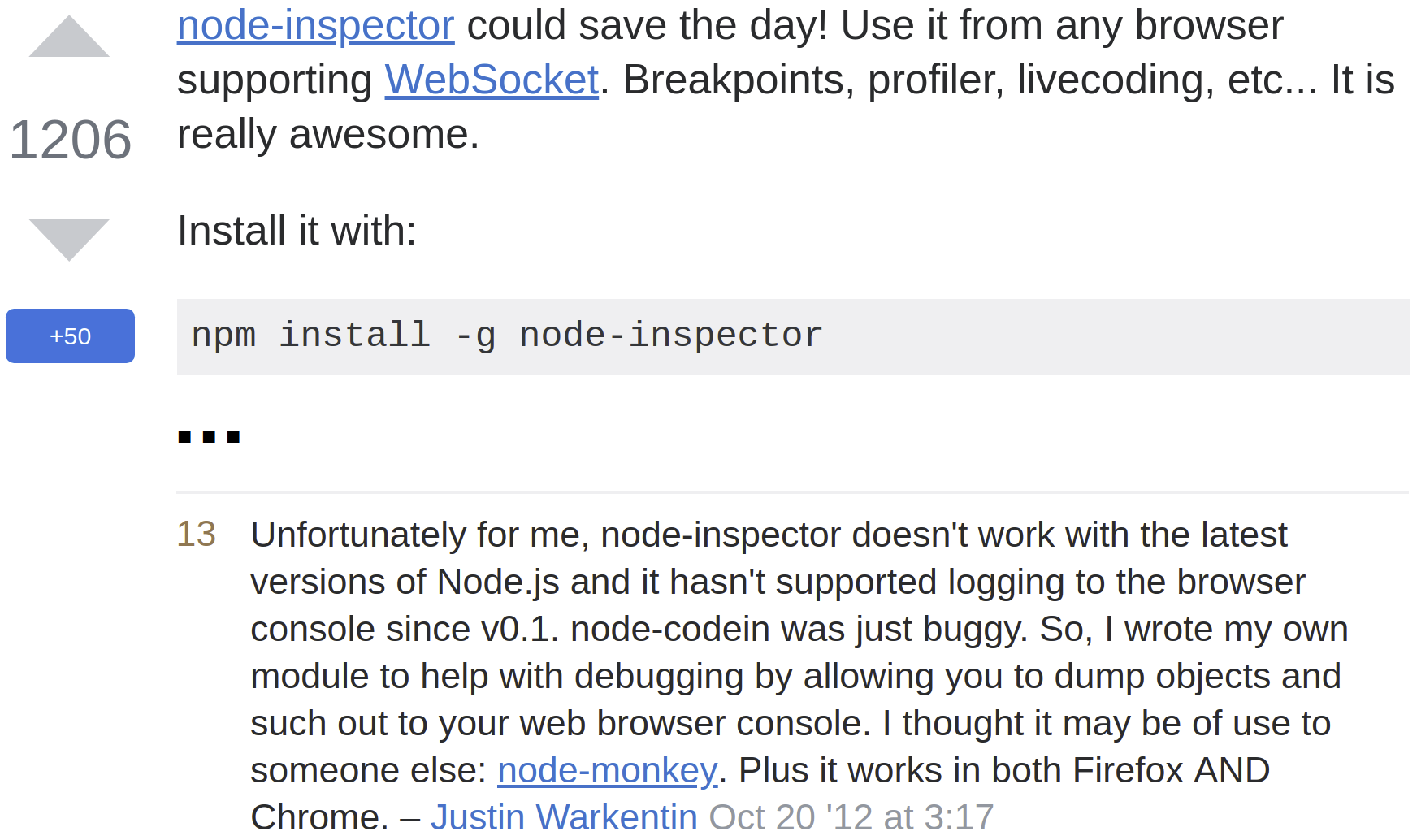}
	\caption{An answer with one of its associated comments. The comment notes that the node-inspector no longer works with the latest versions of Node.js and proposes the use of an alternative module called node-monkey.}
	\label{fig:background_comment_example}
\end{figure}

Commenting on Q\&A websites can lead to more comprehensive discussions which improve the knowledge sharing process~\cite{Gazan:2010,Poche:2017,Zhang:2018}. For example, as shown in Fig.~\ref{fig:background_comment_example}, a user on Stack Overflow posted a comment\footnote{\url{https://stackoverflow.com/posts/comments/17612489}} that pointed out a problem with an existing answer and provided an alternative solution.
On Stack Overflow, comments are ``\textit{temporary `Post-It' notes left on a question or answer}''\footnote{\url{https://stackoverflow.com/help/privileges/comment}}. As of September 2017, users on Stack Overflow have posted 60.2 million comments under questions or answers, which represents more than the total number of questions and answers together according to the data from Stack Exchange Data Explorer\footnote{\url{https://data.stackexchange.com/stackoverflow/query/945995}}. Note that in this study we refer to comments as \emph{the comments that are associated with answers if not specified otherwise} since we focus on studying how comments contribute to providing additional values to answers.

Comments open up a channel for users to add informative discussions to their associated answers. Highly informative comments can expedite the problem solving process, or add knowledge that is worthwhile to share for a particular answer, such as the observation or update of an obsolete answer by comments~\cite{Zhang:2018}. As explained by Jeff Atwood (the co-founder of Stack Overflow), ``\textit{there are often \textbf{important clarifications and addendums left as comments} that \textbf{substantially improve} the original post}''\footnote{\url{https://stackoverflow.blog/2009/04/23/comments-top-n-shown/}}. In addition, Chang et al. note that comments are a \textit{first class citizen} of a Q\&A platform. They can critically improve an answer by providing additional clarifications and refinements to the answer; thus, increasing the overall value of an answer~\cite{Chang:2013}. In Section~\ref{study1.1}, we also observe that comments provide value for various aspects of an answer, such as pointing out errors, making corrections, and providing alternative solutions. Therefore, commenting is an indispensable Stack Overflow feature that leads to improving and ensuring the long-lasting value of crowdsourced knowledge for software developers.

However, uninformative comments could potentially threaten the Stack Overflow community by decreasing the information density of an answer thread (i.e., the compactness of an interface in terms of the amount of information~\cite{Kandogan:1998}). If an increasing number of answer threads are filled up with discussions, such as through uninformative comments, users may exhibit difficulty in identifying relevant information.

Therefore, to promote informative comments and avoid uninformative ones, several sorting rules are applied to comments on Stack Overflow. First, comments can only be posted by the following three types of users: the asker, the answerer, and any user with at least 50 reputation points\footnote{\url{https://stackoverflow.com/help/privileges/comment}}. Stack Overflow also adds a voting system to regulate the quality of comments\footnote{\url{https://meta.stackexchange.com/a/17365}}. Comments can be up-voted, but cannot be down-voted. Thus, the lowest score a comment can have is 0.

In short, commenting is widely used to regulate discussions towards the associated answers. In addition, comments can be as informative as their associated answers to some extent. Thus, in this paper, we study what users actually discuss in comments and characterize the informativeness of comments.

\subsection{Comment hiding mechanism on Stack Overflow}

Initially, Stack Overflow used to hide all comments under answers; however, this rule was abandoned because it hid too much information. As explained by Jeff Atwood that ``\textit{comments were all locked behind ... information was being lost}''\footnote{\url{https://stackoverflow.blog/2009/04/23/comments-top-n-shown/}}. On the other hand, users may find it difficult to locate useful information if all comments were displayed under each answer. The single webpage that contains question, answers, and comments would have an increasing amount of content over time. Hence, users will need to spend more effort and time to read and locate the relevant information.

To have a better balance between displaying all comments and hiding all comments, Stack Overflow has implemented its current comment hiding mechanism to enhance the readability of answer threads.
Since 2009 (Stack Overflow was launched in 2008), this comment hiding mechanism only displays the ``top 5 comments''\footnote{\url{https://stackoverflow.blog/2009/04/23/comments-top-n-shown/}} for each answer (i.e., \textbf{\textit{displayed comments}}). Note that if two comments have the same score (i.e., tied score), they will be ranked by their creation time and the earlier created one will be ranked on the top. Additionally, if more than 30 answers are posted for a question, only the comments with +1 or higher scores will be displayed. In other words, by default, users can read at most 5 displayed comments under any answer. Other comments under an answer are hidden (i.e., \textbf{\textit{hidden comments}}).

\begin{figure}[ht]
	\centering\includegraphics[width=\columnwidth]{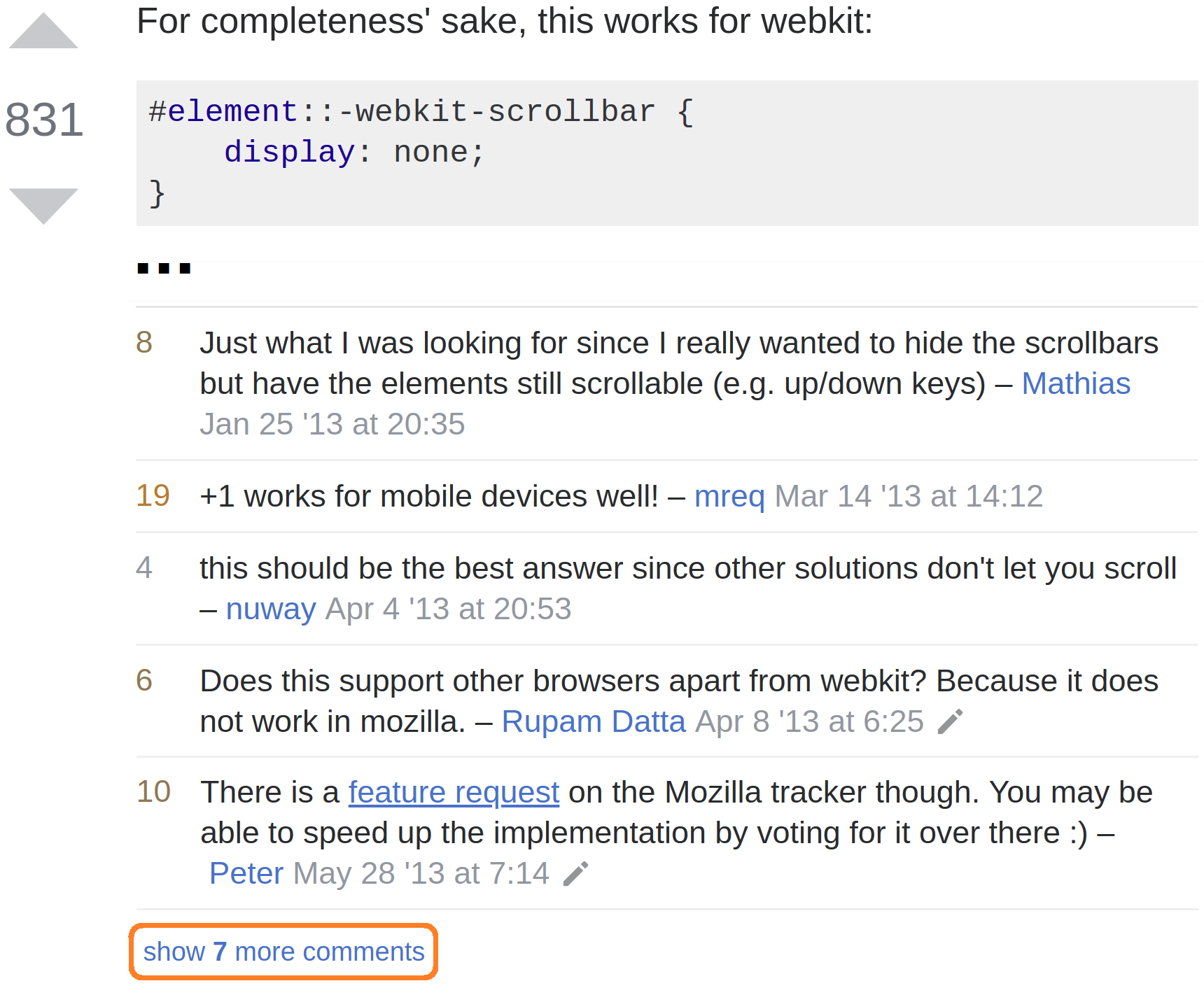}
	\caption{An example of an answer with its displayed comments and a link at bottom saying ``show 7 more comments''. Once a user clicks the link, all comments (including the hidden comments) under the answer will be displayed.}
	\label{fig:hide_comment_example}
\end{figure}

Users can click a link saying ``show n more comments'' to read all comments. An example\footnote{\url{https://stackoverflow.com/a/13184693}} of an answer thread is shown in Fig.~\ref{fig:hide_comment_example} with the displayed comments for the answer and the clickable link. By using the comment hiding mechanism, ideally the comments that do not add additional information to their associated answers are hidden, while the informative comments are displayed.

\begin{table*}[h]
	\centering
	\caption{Statistics of our studied data.}
	\label{table:data_stat}
	\begin{threeparttable} \begin{tabular}{|>{\raggedright}p{2cm}|>{\raggedleft}p{1.1cm}|>{\raggedleft}p{1.2cm}|>{\raggedleft}p{1.6cm}|>{\raggedleft}p{1.3cm}|>{\raggedleft \arraybackslash}p{2.3cm}|}
			\hline
			& \textbf{Number} & \textbf{Proportion} & \textbf{Mean Score} & \textbf{Viewcount*} & \textbf{Mean Viewcount}\\
			\hline
			All Answers & 22,668,556 & 100\% & 2.6 & 2.7e+10 & 2176.7\\
			\hline
			$Answer_{comment}$ & 11,396,766 & 50.3\% & 3.9 & 2.4e+10 & 2873.1\\
			\hline
			$Answer_{noHidden}$ & 10,093,265 & 44.5\% & 3.0 & 2.2e+10 & 2956.0\\
			\hline
			$Answer_{hidden}$ & 1,303,501 & 5.8\% & 11.1 & 8.4e+9 & 6935.4\\
			\hline
		\end{tabular}
		\begin{tablenotes}
			\footnotesize
			\item *We use the viewcount of a question thread to which an answer belongs as the proxy of the viewcount for the answer.
		\end{tablenotes}
	\end{threeparttable}
\end{table*}

The comment hiding mechanism has been in place for nearly 9 years (as of October 2018) while the Stack Overflow community has expanded significantly since the launch of Stack Overflow. The number of users, answers, and associated comments have increased considerably. More and more users are involved in discussions on Stack Overflow through commenting activities~\cite{Zhang:2018}. However, the effectiveness of Stack Overflow's comment hiding mechanism in displaying informative comments and hiding uninformative comments remains unknown. Therefore, we wish to study whether comments (including both hidden and displayed comments) are informative, and we investigate the efficacy of the comment hiding mechanism. Moreover, it is common that developers use search engines to look for solutions to their questions, and many top search results are from Stack Overflow answers. However, hidden comments are not indexed by search engines\footnote{\url{https://stackoverflow.blog/2009/04/23/comments-top-n-shown/}}$^{,}$\footnote{\url{https://meta.stackexchange.com/a/23772}}. As a result, hidden comments get less public attention. It is unknown how the comment hiding mechanism negatively affects developers in locating relevant information. For example, if a user is seeking certain information that is actually from a hidden comment, the comment hiding mechanism would then lead to information loss. In this sense, it is necessary to examine the actual efficacy of the current comment hiding mechanism.

\section{Data collection}
\label{data}
We download the data dump\footnote{\url{https://archive.org/details/stackexchange}} that was published by Stack Exchange in September 2017. We list the statistics of our studied data in Table~\ref{table:data_stat}. There are 22.7 million answers in this dataset. We focus our study on the answers that have comments (i.e., $Answer_{comment}$). 50.3\% (i.e., 11.4 million) of all answers are $Answer_{comment}$,
and there are 32.2 million comments associated with these $Answer_{comment}$. 1.3 million (i.e., 11.4\%) of $Answer_{comment}$ are answers that have comments hidden due to the comment hiding mechanism (i.e., $Answer_{hidden}$). Under such $Answer_{hidden}$, 4.4 million (i.e., 40.5\%) of the comments are hidden. Note that $Answer_{comment}$ includes both $Answer_{hidden}$ and answers that have comments but none of these comments are hidden (i.e., $Answer_{noHidden}$).

In general, $Answer_{hidden}$ are more popular than $Answer_{noHidden}$ on Stack Overflow. More specifically, the median score of $Answer_{hidden}$ is 3.7 times higher than $Answer_{noHidden}$. $Answer_{hidden}$ attract 30.5\% of the viewcounts of all answers across Stack Overflow. The median viewcounts in $Answer_{hidden}$ are 3.2 times larger than the median viewcounts of all answers.


\section{Studying whether comments (including both hidden and displayed comments) are informative}
\label{study1}

\subsection{RQ1: What are the characteristics of both hidden and displayed comments?}
\label{study1.1}

\noindent\textbf{Motivation:} Stack Overflow uses a comment hiding mechanism to split comments under answers into two groups: hidden and displayed comments.
However, it is not clear whether the displayed comments are really more informative than hidden comments as expected. To have a better understanding of hidden and displayed comments, we first conduct a quantitative study to investigate the characteristics of the textual content in both hidden and displayed comments. By knowing this, we can gain insight into the efficacy of the current comment hiding mechanism.\\

\noindent\textbf{Approach:} We study the characteristics of both hidden and displayed comments using a quantitative approach that is described in the following two steps.

We first compare the length of hidden comments with that of displayed comments. The length of a comment is measured by the character number of the comment. We use the length of a comment as a baseline metric to reflect how informative a comment is. In each $Answer_{hidden}$, we calculate the median length ($L_{hidden}$) of all hidden comments within the answer, and the median length ($L_{displayed}$) of all displayed comments within the same answer. The length of the comment is widely used to characterize the quality of comments in other sites (e.g., MetaFilter and YouTube\footnote{\url{http://ignorethecode.net/blog/2009/09/29/comments_size_does_matter/}}). To compare $L_{hidden}$ and $L_{displayed}$, we define the \textbf{\textit{median length ratio of comments}} in a pairwise manner as $Ratio_{L} = L_{hidden} / L_{displayed}$. A value of $Ratio_{L} = 1$ means that the median length of all hidden comments under an answer is equal to the median length of all displayed comments under the same answer. We also compare $L_{hidden}$ with $L_{displayed}$ using the Wilcoxon signed-rank test and the Cliff's delta test~\cite{Cliff:1993} to determine if there is any statistically significant difference between $L_{hidden}$ and $L_{displayed}$.


Second, to understand whether hidden comments add diverse information to the associated answers compared to their displayed comments, We use the vector space model (VSM) to calculate the textual similarity between the hidden comments and the $Answer_{hidden}$ versus the similarity between the displayed comments and the $Answer_{hidden}$. VSM is commonly used for measuring the textual similarity between software engineering artifacts. Readers may refer to the prior studies~\cite{WangLXJ11,WangLL14,ThungWLL13,Chen:2016,Oliveto:2010,Gethers:2011} for more details on VSM.

To apply VSM, we treat each answer as one document, all of its associated hidden comments together as one document, and all of its associated displayed comments together as one document. For each document, we first perform the following common pre-processing steps~\cite{WangLL14,Chen:2016}: remove HTML tags/URL, split words by punctuation marks, split words using camel cases, convert upper case letters to lower case letters, and remove stop words. We then convert each pre-processed document to a vector, in which the weight of each element of the vector is calculated based on term frequency (i.e., the frequency of the term in the document) and inverse document frequency (i.e., the reciprocal of the number
of documents containing the term). Finally, we compute the cosine similarity between answers and their associated comments (hidden and displayed, respectively).

In order to compare the textual similarity between hidden and displayed comments under the same answer, we calculate the cosine similarity in a pairwise manner. In each $Answer_{hidden}$, we calculate the cosine similarity ($S_{Answer~vs.~Hidden}$) between the answer and all of its associated hidden comments. In the same $Answer_{hidden}$, we calculate the cosine similarity ($S_{Answer~vs.~Displayed}$) between the answer and all of its associated displayed comments. We define the \textbf{\textit{pairwise cosine similarity ratio}} as $Ratio_{S} = S_{Answer~vs.~Displayed}$ / $S_{Answer~vs.~Hidden}$. Note that a value of $Ratio_{S}~\textgreater{}~1$ means that, the displayed comments are more similar to their associated answer as compared to that of the hidden comments.\\

\begin{figure}[ht]
\centering\includegraphics[width=\columnwidth]{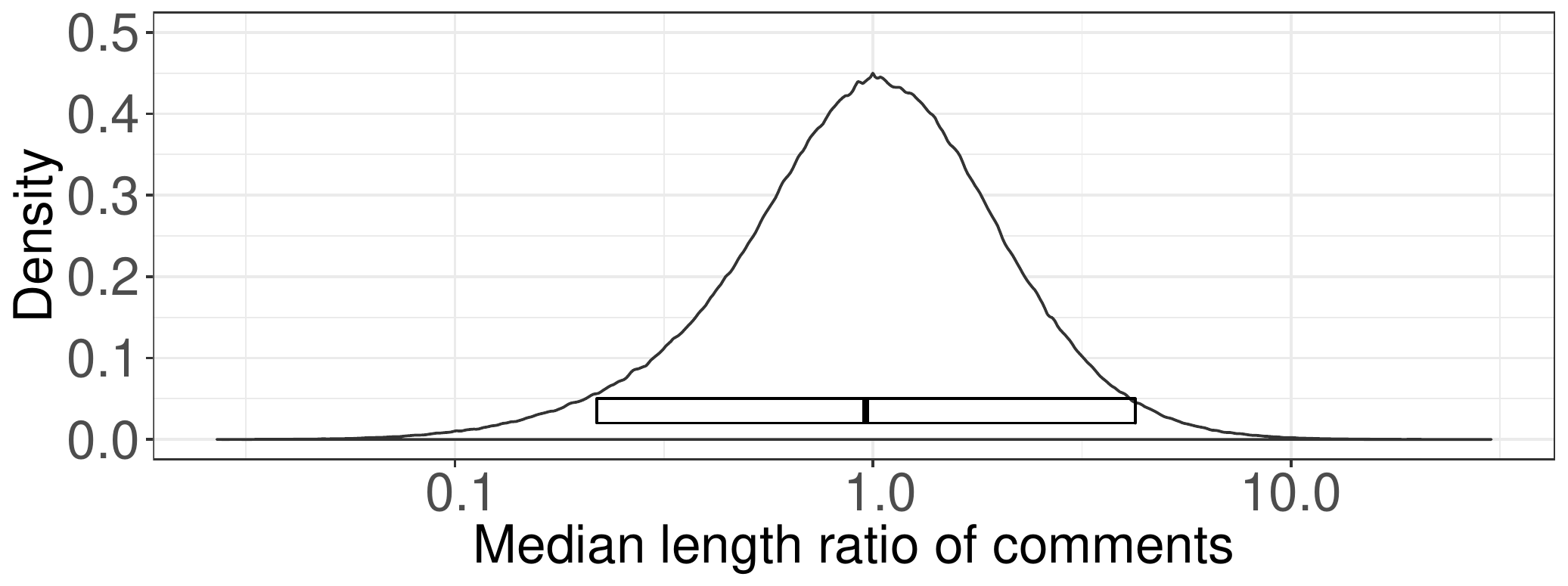}
\caption{The distribution of the median length ratio of comments in a pairwise manner ($Ratio_{L}$). $Ratio_{L}$ = 1 means that the median length of all hidden comments is equal to the median length of all displayed comments under the same answer.}
\label{fig:median_char_length}
\end{figure}

\noindent\textbf{Results:} \textbf{There is no statistically significant difference between hidden and displayed comments in terms of median length.} We plot the distribution of $Ratio_{L}$ in $Answer_{hidden}$ (as shown in Fig.~\ref{fig:median_char_length}). $Ratio_{L}$ shows a normal distribution with a mean value of 1. Our statistical test results show that there is no statistically significant difference between hidden and displayed comments in terms of length (p-value $>$ 0.05). 

\begin{figure}[ht]
    \centering\includegraphics[width=\columnwidth]{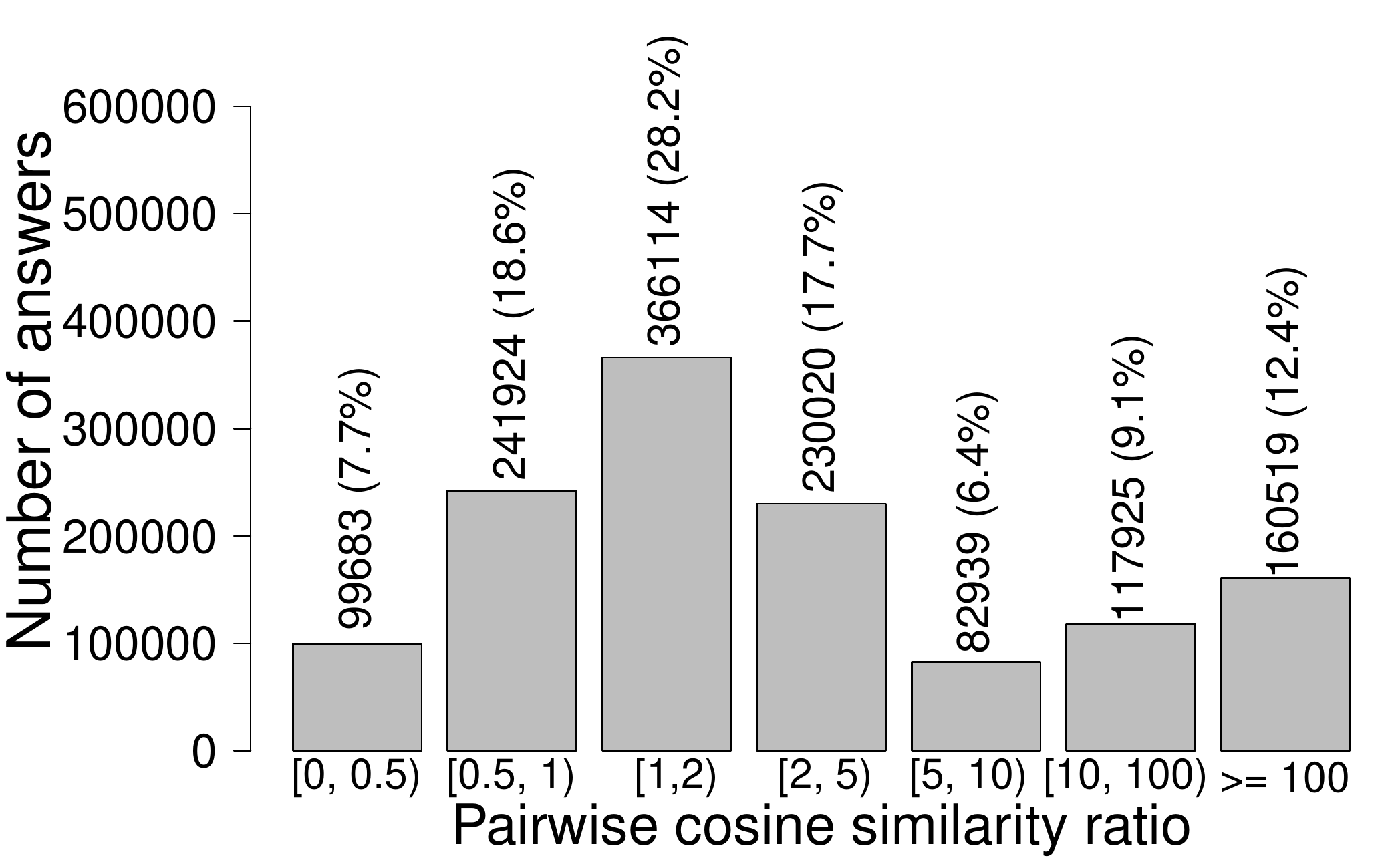}
    \caption{The distribution of the pairwise cosine similarity ratio $Ratio_{S}$. A higher $Ratio_{S}$ indicates that displayed comments are more similar to their associated answer as compared with the hidden comments to the same answer.}
    \label{fig:cosine_similarity}
\end{figure}

\textbf{In the majority of $Answer_{hidden}$, hidden comments share less semantic similarity with the associated answers than that of displayed comments. The finding may suggest that even though the hidden comments have the same amount of text compared with the displayed comments, the hidden ones have a richer vocabulary and add a greater variety of content than the displayed comments to the associated answers.} The distribution of $Ratio_{S}$ is shown in Fig.~\ref{fig:cosine_similarity}. In 73.8\% of $Answer_{hidden}$, the cosine similarity between the displayed comments and the associated answers are no less than the cosine similarity between the hidden comments and the associated answers. The Wilcoxon signed-rank test shows a statistically significant difference (p-value $<$ 0.05) for the cosine similarity between displayed comments to their associated answers and hidden comments to their associated answers. The Cliff's delta test also shows that the difference is medium (i.e., -0.39).

\rqbox{
    Hidden comments have the same amount of text as displayed comments. Hidden comments have a richer vocabulary and add a greater variety of textual content to their associated answers than that of displayed comments to the same answer.
}

\vspace{-0.15in}

\subsection{RQ2: What do users discuss in hidden and displayed comments?}
\label{study1.2}

\noindent\textbf{Motivation:} As shown in Section~\ref{background}, comments can be informative, thus augmenting their associated answers with valuable knowledge. Identifying what users discuss in comments helps us better understand how comments actually augment their associated answers; providing insights for improving the current commenting system.\\

\noindent\textbf{Approach:} To capture the information that users could obtain from comments, we investigate what users discuss in both hidden and displayed comments. We perform a qualitative study to understand what is discussed in hidden and displayed comments. To do so, we randomly select 384 hidden comments from $Answer_{hidden}$, and randomly select 384 displayed comments from $Answer_{hidden}$, in order to obtain a statistically representative sample with a 95\% confidence level and a 5\% confidence interval~\cite{Boslaugh:2012}.

We manually label the category of discussion in each comment. If a comment has multiple sentences, we assign a label for each sentence individually. Therefore, one comment can be assigned with multiple labels because a user may discuss more than one topic in a comment. We perform a lightweight open coding-like process that is similar to Seaman et al.~\cite{seaman1999qualitative, seaman2008defect, Zhang:2018} to identify the category of topics that are discussed in a comment (i.e., \emph{comment category}). This process involves 3 phases and is performed by the first two authors (i.e., A1--A2) in this paper:

\begin{itemize}
    \item Phase I: A1 identifies a draft list of the categories of comments based on a sample of 50 comments from hidden comments and another 50 comments from displayed comments. Then, A1 and A2 use the draft list to label the comments collaboratively, during which the categories are revised and refined.

    \item Phase II: A1 and A2 independently apply the resulting categories from Phase I to label all 768 comments (i.e., 384 hidden comments and 384 displayed comments). A1 and A2 take notes regarding the deficiency or ambiguity of the already-identified categories for labeling certain comments. Note that new categories are added during this phase if A1 and A2 observe the need for more categories. At the end of this phase, we end up with 7 categories of comments (see Table~\ref{tab:comment_type}). Cohen's kappa~\cite{Gwet:2002} is used to measure the inter-rater agreement, and the kappa value is 0.72 (measured at the end of Phase II), implying a high level of agreement.

    \item Phase III: A1 and A2 discuss the coding results obtained in Phase II to revolve any disagreement until a consensus is reached. No new categories are added during this discussion.
\end{itemize}

\begin{table}[ht]
	\centering
	\caption{Comment categories}
	\label{tab:comment_type}
	\begin{tabular}{|p{2cm}|p{6cm}|}
		\hline
		\textbf{Category} & \textbf{Explanation}\\
		\hline
		\textbf{Praise} & Praise an answer\\
		\hline
		\textbf{Advantage} & Discuss the advantage of an answer\\
		\hline
		\textbf{Improvement} & Make improvement to an answer\\
		\hline
		\textbf{Weakness} & Point out the weakness of an answer\\
		\hline
		\textbf{Inquiry} & Make inquiry based on an answer\\
		\hline
		\textbf{Addition} & Provide additional information to an answer\\
		\hline
		\textbf{Irrelevant} & Discuss irrelevant topics to an answer\\
		\hline
	\end{tabular}
\end{table}

We also compare if there is a statistically significant difference in the categories of hidden and displayed comments using Wilcoxon signed-rank test and Cliff's delta test.\\

\noindent\textbf{Results:} \textbf{More than half of the comments are informative in both hidden and displayed comments.} Except for category \textit{praise} and \textit{irrelevant}, we consider the comments of other categories (i.e., \textit{advantage}, \textit{improvement}, \textit{weakness}, \textit{inquiry}, and \textit{addition}) as \textbf{\textit{informative}}. Note that we consider the category \textit{praise} as redundant for up-voting answers, thus we do not considered such comments as informative. The category \textit{irrelevant} also does not add any direct value to the question answering process.

The distribution of comment categories \textit{advantage}, \textit{weakness}, \textit{inquiry} and \textit{addition} are very similar between hidden and displayed comments. The top category is \textit{addition} in both hidden and displayed comments. In this category, users provide additional information to the associated answers of the comment. Namely, by providing an alternative answer to a question, adding an example, adding an explanation, or adding a reference. An example of an informative comment is shown in Fig.~\ref{fig:background_comment_example}, in which a user pointed out that the node-inspector did not work any more in the latest Node.js version (category \textit{weakness}), he also provides an alternative in the same comment (category \textit{addition}).

\begin{figure}[ht]
    \centering\includegraphics[width=\columnwidth]{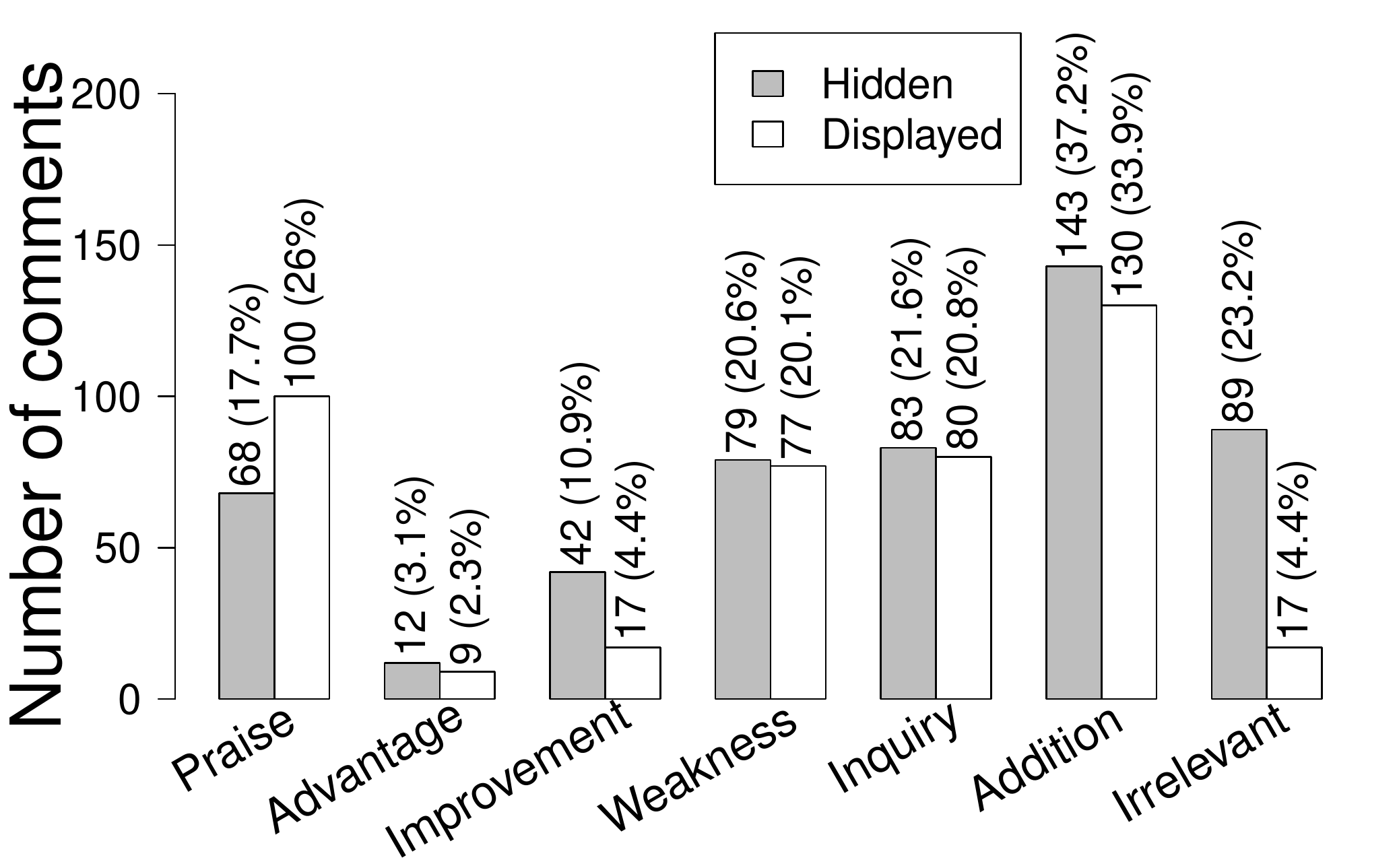}
    \caption{The distribution of comment categories.}
    \label{fig:comment_type}
\end{figure}

\textbf{There is no statistically significant difference between hidden and displayed comments in terms of the distribution of comment categories.} Fig.~\ref{fig:comment_type} shows the distribution of comment categories for both hidden and displayed comments. The result of the Wilcoxon signed-rank test shows the differences between hidden and displayed comments are insignificant (i.e., p-value $>$ 0.05). The result of Cliff's delta test is negligible (i.e., 0.14). The comparison of the proportion of informative comments in both hidden and displayed comments is shown in Table~\ref{table:comment_informative}. The studied hidden comments share similar information (in terms of the comment categories) compared to displayed comments. That being said, \textbf{hidden comments are as informative as displayed comments for the associated answers}.

We also note that commenters with higher reputation points are more likely to post more informative comments. In the studied hidden comments, the median reputation points of these commenters are 595.5 and 457.5 for the informative and uninformative comments respectively. In the studied displayed comments, the median reputation points of these commenters are 677 and 364 for the informative and uninformative comments respectively.\\

\begin{table}[ht]
    \centering
    \caption{Comparison of hidden and displayed comments in providing additional value to the associated answers}
    \label{table:comment_informative}
    \begin{tabular}{|p{2.6cm}|p{1.8cm}|p{2.2cm}|p{0.5cm}|}
        \hline
        & \textbf{Informative} & \textbf{Not informative} & \textbf{Total}\\
        \hline
        \textbf{Hidden comment} & 280 (72.9\%) & 104 (27.1\%) & 384\\
        \hline
        \textbf{Displayed comment} & 289 (75.3\%) & 95 (24.7\%) & 384\\
        \hline
    \end{tabular}
\end{table}

\rqbox{
    Hidden comments are as informative as displayed comments. More than half of the comments are informative in both hidden and displayed comments.
}

\vspace{-0.15in}

\section{Studying the efficacy of the comment hiding mechanism}
\label{study2}

In Section~\ref{study1}, we discover that the majority of the hidden comments are at least as informative as the displayed comments. Thus, although the comment hiding mechanism has been designed to hide uninformative comments, many informative comments are also hidden. 
In addition, developers commonly make use of web search engines, such as Google, to locate online resources to improve their productivity~\cite{Xia2017}. However, Google does not index such a large number of hidden comments instead it only indexes the displayed comments, which prevents users from accessing the information in these hidden comments from search engines. Therefore, we focus on studying the current comment hiding mechanism, that is, the principle that determines whether a comment should be hidden or displayed. By investigating the efficacy of the current comment hiding mechanism, we wish to offer deeper insights into enhancing the Stack Overflow commenting system, so that users can more conveniently and effectively perceive informative discussions through comments.

\subsection{RQ3: How effective is the comment hiding mechanism?}
\label{study2.1}
\textbf{Motivation:}
Stack Overflow's current comment hiding mechanism aims at displaying comments with higher scores while hiding ones with lower scores. The assumption is that comments with higher scores are more informative than the ones with lower scores.
However, in Section~\ref{study1} we find that hidden comments are as informative as displayed comments, which suggests that the current comment hiding mechanism is not working as expected.
Therefore, it is important to investigate the reason behind this.

\begin{figure}[ht]
	\centering\includegraphics[width=\columnwidth]{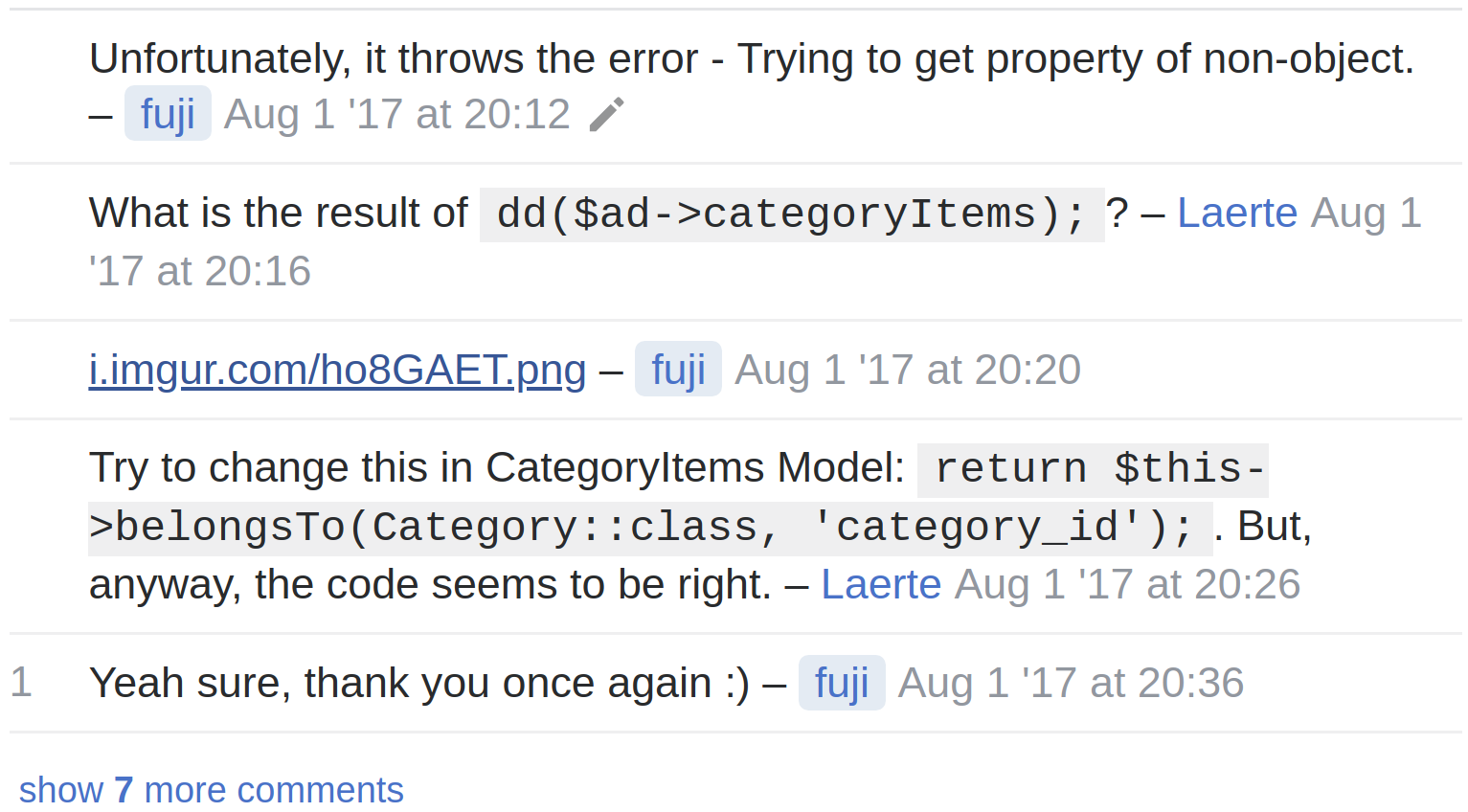}
	\caption{An example of an answer where a large proportion (i.e., 11) of the comments under an answer have 0 score and only 1 comment has a score of 1.}
	\label{fig:example_many_zero}
\end{figure}

During our manual study, we also notice that this comment hiding mechanism may not work well if comments do not have a hierarchy of different scores. For example, if many comments do not get any up-vote (i.e., their scores are 0), apparently, the current comment hiding mechanism would not work in such a scenario. More generally, as long as comments have the same score, the comments would not be ranked nor displayed based on their scores (i.e., tie-scored comments). As an example, in an answer\footnote{\url{https://stackoverflow.com/a/45446651}}, there are 12 comments with only 1 comment with a non-zero score (i.e., 1) as shown in Fig.~\ref{fig:example_many_zero}. In such cases, the displayed comments may not be more informative than the hidden comments.

In order to study the efficacy of the comment hiding mechanism, in this RQ, we investigate how comments are actually ranked and therefore displayed.

\noindent\textbf{Approach:} Intuitively, an uninformative comment (i.e., category \textit{Praise} and \textit{Irrelevant}) should be hidden by the comment hiding mechanism so that another informative comment (such as category \textit{Improvement} and \textit{Weakness}) could be displayed under the same answer. The comment hiding mechanism is designed for this purpose, i.e., re-arrangement of comments based on their scores. To evaluate the efficacy of the comment hiding mechanism in action, we first characterize the comment scores and analyze how they affect the comment hiding mechanism, since the current comment hiding mechanism is designed based on the comment score.

\begin{figure}[ht]
	\centering\includegraphics[width=\columnwidth]{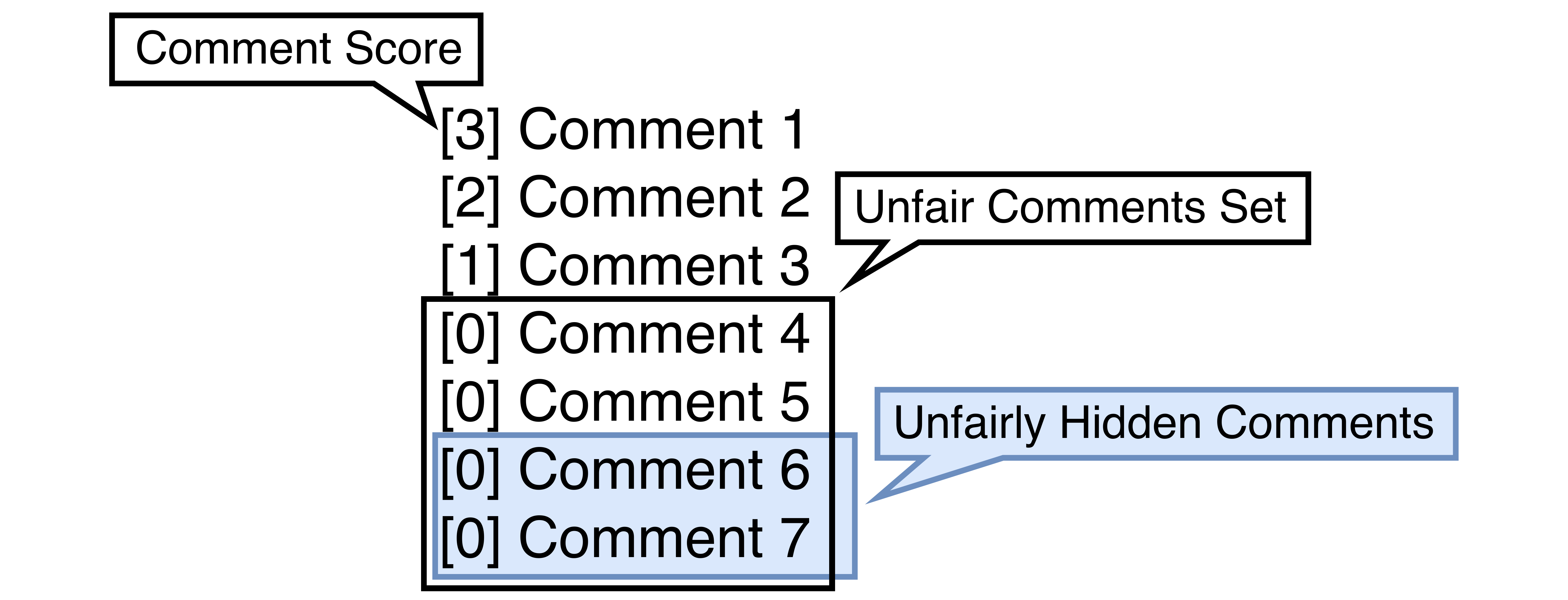}
	\caption{An example of an unfair comments set and its unfairly hidden comments. Comment 6 and 7 are unfairly hidden comments since they have the same score as Comment 4 and 5; however, Comment 6 and 7 are hidden.}
	\label{fig:unfair_comment_set}
\end{figure}

More specially, we investigate how the tie-scored comments impact the comment hiding mechanism. For this purpose, we make the following definitions. We define that a comment is \textbf{\textit{unfairly hidden}} when it is hidden not because it has a lower score than another comment, but because it is posted later than another displayed comment with the same score (i.e., \textit{unfairly hidden comments}). In other words, an unfairly hidden comment is hidden because of its later creation time instead of its lower score (note that all such unfairly hidden cases only happen in tie-scored comments based our definition). We show an example of unfairly hidden comments and unfair comments set in Fig.~\ref{fig:unfair_comment_set}. Currently, an unfairly hidden comment occurs in the following situation: for all comments of an answer sorted by score, the score of the sixth comment (i.e., a hidden comment) is equal to the score of the fifth comment (i.e., a displayed comment). In this situation, the fifth comment does not need to compete with the sixth comment to be displayed by the user interface, it gains its position (as a displayed comment) because it is created earlier. 

Furthermore, we define a set of comments under an answer as an \textbf{\textit{unfair comments set}}, if there are some hidden and displayed comments that have the same score (e.g., a comment with a score of 0 is hidden but another comment with a score of 0 is displayed, see Fig.~\ref{fig:unfair_comment_set}). We conduct a quantitative study to find out how many answers have unfairly hidden comments. If there were a large proportion of such unfair comments set, it may indicate that the current comment hiding mechanism is not working as expected.



Besides the above-mentioned aspects related to comment score, we calculate the proportion of $Answer_{hidden}$ in which their comments are actually ordered and displayed by their creation time (i.e., the comment ordering mechanism has no effect). By investigating these characteristics, we wish to understand the impact of comment scores and creation times on the current comment hiding mechanism.\\


\noindent\textbf{Results: Due to the widespread existence of tie-scored comments, unfairly hidden comments exist in 97.3\% (i.e., 1,268,416 out of 1,303,501) of the $Answer_{hidden}$.} Currently, the comment hiding mechanism fails to consider tie-scored comments, leading to new comments being hidden while old comments with the same score being displayed in almost all $Answer_{hidden}$ (i.e., resulting in the stagnation of displaying new comments). Even more, unfairly hidden comments sets have 4,105,956 hidden comments. In other words, 92.9\% of all the 4,418,563 hidden comments are actually unfairly hidden (i.e., they are hidden not because of the score or content but the time they are posted). To illustrate the issue, in the same example shown in Fig.~\ref{fig:example_many_zero}, only 1 comment has a score of 1 while the other 11 comments have a score of 0; therefore, 4 of the 0-score comments are displayed simply because they were created earlier than the other 7 comments with 0-score. Therefore, any new comment, even if they are informative, will be automatically hidden. The lack of visibility makes unfairly hidden comments less likely to get any up-voting, and thus are even more likely to remain hidden. 

\textbf{944,950 (i.e., 72.5\%) of $Answer_{hidden}$ have unfairly hidden comments with a score of 0. More than half (i.e., 56.5\%) of $Answer_{hidden}$ have all of their comments with the same score of 0.} In other words, as an upper bound estimation, the comment hiding mechanism surprisingly only works as expected at most for 43.5\% of $Answer_{hidden}$. Moreover, even in such 43.5\% cases, it is not guaranteed that every single comment is ranked based on the score since perhaps only a portion of comments have a score greater than 0. For example, in Fig.~\ref{fig:example_many_zero}, the answer has 12 comments, and only one of them has a score greater than 0 while the remaining of comments have a score of 0. In this example, the remaining 11 comments are not ranked based on their score anymore. We notice that 87.7\% of all the comments under all answers have a score of 0. One possible reason for such a large number of comments that do not have any up-voting is as Calefato et al. mentioned in their previous study, comments are considered as a ``free zone'' for users since comments do not generate any reputation point~\cite{Calefato:2015}. Thus, users may not be motivated to up-vote comments.

Comments are ranked based on their creation time if their scores are the same. Given the fact that most of $Answer_{hidden}$ have all of their associated comments with the same score, we wish to determine how many $Answer_{hidden}$ have their associated comments actually ordered by the comment creation time.

\textbf{In 79.4\% of $Answer_{hidden}$, comments are ranked and displayed by the order of their creation time.} In these answers, the result of the comment hiding mechanism is equivalent to a queue of comments that are sorted by the creation time of their comments. Namely, only the first 5 oldest comments are displayed, and any newer comment will be hidden. In other words, \textbf{the current comment hiding mechanism gives priority to older comments --- promoting stagnation of comments}. As we explained before, one possible reason that a large proportion of $Answer_{hidden}$ are actually displayed based on their creation time is the widespread existence of 0-score comments.

\textbf{Another possible explanation is that older comments tend to get higher scores.} Note that the comment age is defined as the time interval between the creation of the comment and its associated answer. Among the 303,035 $Answer_{hidden}$ that have at least 2 comments whose scores are \textgreater= 1, 187,714 (i.e., 61.9\%) have a negative correlation, and 65,205 (21.5\%) have at least a moderate negative correlation (correlation \textless~-0.5)~\cite{mukaka2012guide} between comment age and score. Among the 46,935 $Answer_{hidden}$ that have at least 5 comments whose scores are \textgreater= 1, 36,655 (i.e., 78.1\%) have a negative correlation, and 15,525 (i.e., 33.1\%) have at least a moderate negative correlation (correlation \textless~-0.5) between comment age and score. Therefore, older comments are more likely to get higher scores.



\rqbox{
    The current comment hiding mechanism does not work effectively. If an answer has hidden comments, it is highly likely (97.3\%) that it has unfairly hidden comments. The current mechanism fails to consider the widespread of comments with tie-score, especially 0-score, and gives a higher priority to displaying older comments.
}

\vspace{-0.15in}

\subsection{RQ4: What are the characteristics of unfairly hidden comments?}
\label{study2.2}
\textbf{Motivation:} In Section~\ref{study2.1}, by examining the age of commentd, their score, and their correlation, we find that, in most cases, the current comment hiding mechanism actually fails to rank and display comments based on their scores. The current mechanism does not consider tie-scored comments (i.e., comments that have the same score). For example, during our manual study in Section~\ref{study2.2}, we observe that some unfairly displayed comments are very short and uninformative (e.g., expressing praise) while some unfairly hidden comments are informative. In order to improve the current comment hiding mechanism, we investigate the characteristics of displayed and hidden comments in the unfair comments set. By understanding this, we can provide insightful suggestions for improving the current comment hiding mechanism for Stack Overflow.\\



\noindent\textbf{Approach:} 
We first investigate the length of unfairly displayed comments as a baseline metric to measure how informative they are. If some unfairly comments are very short, it is highly likely that they are uninformative. Furthermore, we investigate the informativeness of the shortest unfairly displayed comment compared with the longest unfairly hidden comment in the same unfair comments set. The reason that we conduct such comparison is because we probably could provide insights into improving the comment hiding mechanism. For example, one simple solution is to replace the shortest unfairly displayed comment with the longest unfairly hidden comment if we can show that the longest unfairly hidden comment is more likely to be informative than the shortest unfairly displayed comment in the same unfair comments set. To do so, we randomly select 384 sets of comments in the unfair comments sets, with at least 1 unfairly displayed comment with length \textless{} 50 and 1 unfairly hidden comment with length \textgreater{}= 50 to achieve a significance level of 95\% and a significance interval of 5\%. We manually label the comment category using the same qualitative approach in Section~\ref{study1.1}, for both the shortest unfairly displayed comment and the longest unfairly hidden comment in each sampled set of comments. The Cohen's Kappa value is 0.81 before discussion.

We then perform a qualitative analysis to investigate the comment categories (see Table~\ref{tab:comment_type}) of each unfairly hidden and unfairly displayed comment pair to see if an unfairly hidden comment would be more informative than the corresponding unfairly displayed comment under the same answer.\\

\begin{figure}[ht]
	\centering\includegraphics[width=\columnwidth]{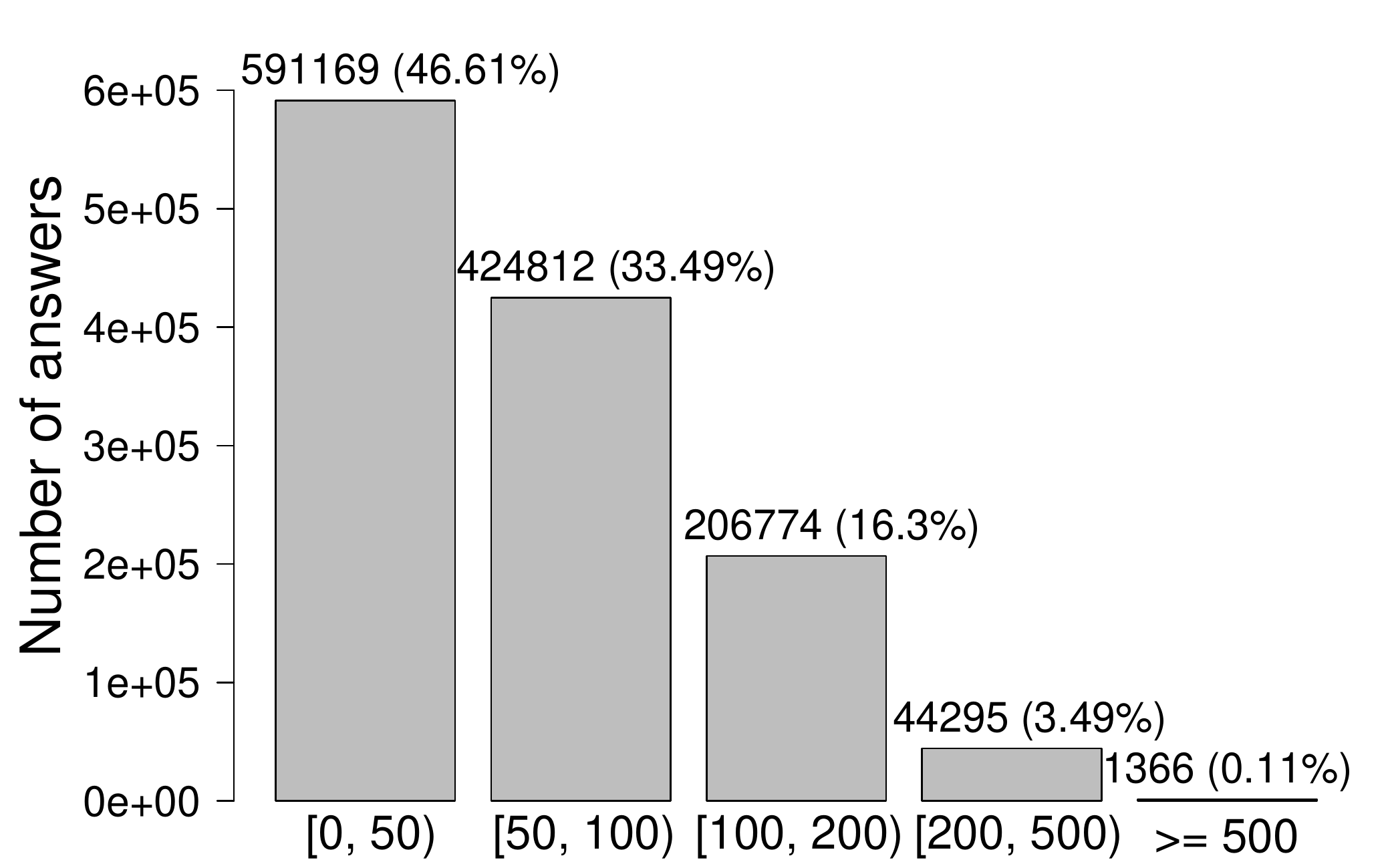}
	\caption{The distribution of the shortest unfairly displayed comments.}
	\label{fig:shortest_unfairly_displayed_comment}
\end{figure}

\noindent\textbf{Results:} \textbf{In around half (i.e., 46.6\%) of the answers that have unfairly hidden comments, the shortest unfairly displayed comments have a length that is less than 50 characters.} Fig.~\ref{fig:shortest_unfairly_displayed_comment} shows the distribution of answers that have unfairly displayed comments against different ranges of the length of the shortest comment under the same answers. 
Through our observation, we find that short comments are usually not informative. For example, a short comment saying ``of course ... that's obvious''\footnote{\url{https://stackoverflow.com/posts/comments/56897172}} does not add any information to the associated answer.

\begin{figure}[ht]
	\centering\includegraphics[width=\columnwidth]{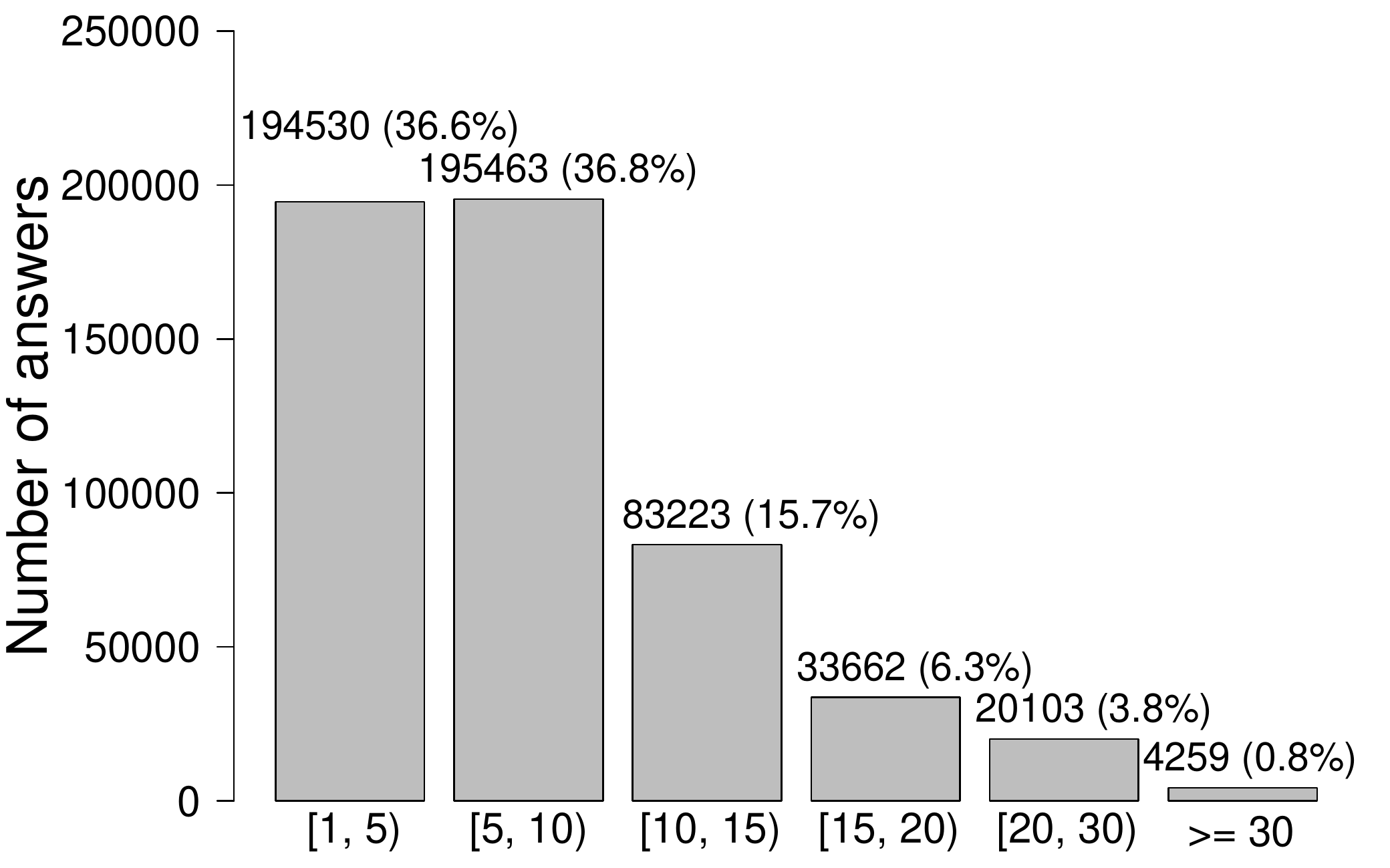}
	\caption{The distribution of the length ratio.}
	\label{fig:length_ratio}
\end{figure}

More specifically, in answers that have unfairly hidden comments with the shortest unfairly displayed comments being less than 50 characters ($L_{DisplayedMin}$), we pick the longest unfairly hidden comment ($L_{HiddenMax}$) in the same unfair comments set, and calculate the length ratio as $Ratio_{unfair} = L_{HiddenMax} / L_{DisplayedMin}$. The distribution of $Ratio_{unfair}$ is shown in Fig.~\ref{fig:length_ratio}. \textbf{In 63.4\% of such cases, the length of the longest unfairly hidden comment is at least 5 times as long as the length of the shortest unfairly displayed comment.} Such a high ratio between the longest unfairly hidden comment and the shortest unfairly displayed comment of the same answers may indicate that the longest unfairly hidden comment is more informative than the shortest one.

\begin{figure}[ht]
\centering\includegraphics[width=\columnwidth]{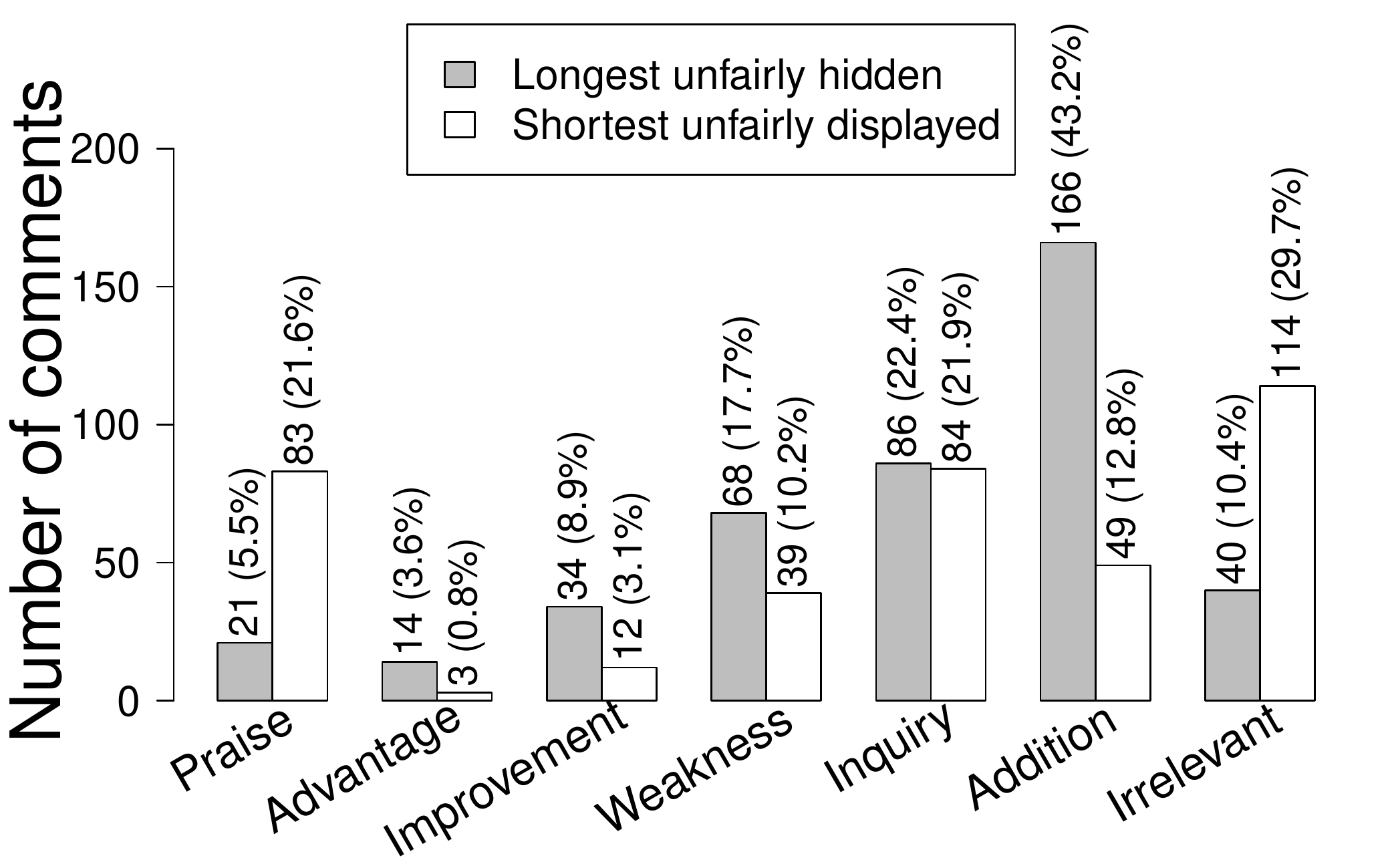}
\caption{The distribution of the categories of comments in both the shortest unfairly displayed and the longest unfairly hidden comments.}
\label{fig:manual_compare}
\end{figure}

\textbf{In cases where the shortest unfairly displayed comment has fewer than 50 characters in the unfairly hidden comments set, the longest unfairly hidden comment is more likely to be informative than the shortest unfairly displayed comment.} As shown in Fig.~\ref{fig:manual_compare}, in the unfairly hidden comments, only 15.9\% of comments are related to irrelevant information and praise, while in unfairly displayed comments 51.3\% are related to irrelevant information and praise. As a result, Stack Overflow could replace such short displayed comments with another long hidden comment from the unfair comments set.

\noindent\textbf{Discussion:} As an exploratory experiment, we inspect how the comment hiding mechanism impacts certain informative observations in comments, such as answer obsolescence~\cite{Zhang:2018}, security vulnerability, and error message. An example of such a comment\footnote{\url{https://stackoverflow.com/posts/comments/78752711}} says ``\textit{works awesome, the only thing that i had to change was the onAttach of the fragment, since it has been deprecated}''. This comment pointed out that the onAttach() function of the Android Fragment class is deprecated; however, this comment is hidden by the comment hiding mechanism while none of the currently displayed comments bring up this deprecation issue. This motivates us to search among all $Answer_{hidden}$ for comments that mention the word ``obsolete'' or ``outdate'', and we find that in such 6,523 comments of answer obsolescence observations, 42.5\% are actually hidden. \textbf{Furthermore, 85.5\% (i.e., 2,370) of the 2,771 hidden comments of answer obsolescence observations are actually unfairly hidden by the comment hiding mechanism.} Since software obsolescence are more likely to happen over time, the observation of answer obsolescence tends to happen in newer comments instead of older ones. In Section~\ref{study2.1}, we find that in the majority of $Answer_{hidden}$, comments are ranked and displayed by their creation time. Therefore, the current comment hiding mechanism is much more likely to hide comments that observe answer obsolescence. If these unfairly hidden observations of answer obsolescence could have been replaced by other unfairly displayed comments, users would be more aware of the answer obsolescence issue on Stack Overflow.

In addition, we find similar trends from other informative observations in comments. For example, among comments mentioning the word ``vulnerable'' (i.e., 1,603), 38.9\% are hidden. Among comments mentioning the word ``error'' (i.e., 566,756), 36.7\% are also hidden.

Among all of the above-mentioned observations that are related to answer obsolescence, security vulnerability, or error message, a significant proportion of such observations are buried in hidden comments. Our finding suggests that the comment hiding mechanism does not facilitate users in finding these obvious flaws in answers. Note that in Section~\ref{study2.1}, we find that the comment hiding mechanism only applies to less than half of $Answer_{hidden}$ in the best-case scenario while in the remaining of $Answer_{hidden}$ all comments (due to all of them having a 0-score) are simply ranked and displayed by time. Therefore, the comment hiding mechanism may bury other informative observations as well.

\rqbox{
	In the unfair comments set, the longest unfairly hidden comment is more likely to be informative than the shortest unfairly displayed comment, especially if the shortest unfairly displayed comment has fewer than 50 characters. As a solution to improve the comment hiding mechanism, Stack Overflow can swap these pairs of comments.
}

\vspace{-0.15in}

\section{Discussion}
\label{discussion}
\textbf{Stack Overflow should improve their current comment hiding mechanism by considering the tie-scored comment cases.} The comment hiding mechanism is unable to prioritize any comment among tie-scored comments. Therefore, these comments could simply be ranked by their creation time, and any new comment other than the oldest 5 can be hidden. Older comments are more likely to be displayed and get attention (e.g., have a higher score), while newer comments are more likely to be hidden. The observation exhibits the \textbf{\textit{Matthew effect: the rich get richer and the poor get poorer}}~\cite{Merton:1956}. One possible approach to alleviate the issue from the comment hiding mechanism is to replace shorter displayed comments with longer hidden ones. Another possibility is to randomly display comments in the unfair comments set. Other algorithms focusing on sorting tie-scored unfair comments set can be exploited to effectively hide noisy comments while still retaining informative comments\footnote{\url{https://meta.stackexchange.com/q/204402}}. Stack Overflow can also allow for the down-voting of comments to break the tie-score. This effort can continue to improve the overall quality of knowledge sharing on Stack Overflow.

Another possible solution is to develop an automated classifier to identify informative comments from uninformative comments in the unfair comments set. Although Stack Overflow allow users to up-vote comments, and comments with a higher score are more likely to be among the top 5 thus displayed, the current comment hiding mechanism does not effectively reinforce its goal. By using this automated approach, the mechanism could be optimized without massive effort of manual labeling by users. It could also assist users in flagging comments (i.e., as a way of bringing inappropriate content to the attention of the community\footnote{\url{https://stackoverflow.com/help/privileges/flag-posts}}, such as labeling unfriendly or unkind comments). Moderators process approximately 1500 flags per day\footnote{\url{https://meta.stackexchange.com/a/166628}}. The classifier could indicate whether a comment is informative from the flagged comments, so he/she can efficiently determine noisy comments for removal and informative comments to keep.

\textbf{Users are encouraged to read through all comments (including hidden comments) in case any further corrections are made in such comments, such as observations of answer obsolescence, security vulnerability, and error message.} The value of Stack Overflow answers can change over time even when these answerers have not yet noticed the change. Therefore, comments provide another channel to notify a wider audience on Stack Overflow about any change to the existing answer. Especially in highly attractive answers, many comments are hidden without taking into account whether they are informative observations to answers. To prevent using an obsolete solution, an insecure code snippet, or a running error, users are encouraged to read through all comments under an answer before attempting to solve their issues based on the answer, especially more recent ones because they are more likely to be hidden.

\section{Threats to Validity}
\label{validation}
\textbf{External validity:} Threats to external validity are related to the generalization from our study. In this study, we focus on the comment hiding mechanism of Stack Overflow, which is the most popular technical Q\&A sites in the world. However, our findings and suggestions may not generalize well to other Q\&A sites (especially, other sites under Stack Exchange that have the same question, answer, comment layout). 
Future studies could focus on other Q\&A sites since some of these sites (such as Super User, Server Fault, and Ask Ubuntu) contribute significantly to knowledge sharing in their specific domains.

We conduct two qualitative studies in our case study, the first of which investigates what users discuss in both hidden and displayed comments, and the second one explores whether longer unfairly hidden comments are more informative than shorter unfairly displayed comments. Since it is impossible for us to manually study all comments in this study, we attempt to minimize the bias by selecting a statistically representative samples of comments with a 95\% confidence level and a 5\% confidence interval.

\noindent\textbf{Internal validity:} Threats to internal validity are related to experimental errors. Comment categories are determined by the authors of this study, and later on the informative comment categories are evaluated by the same authors. To reduce the bias of this process, each comment is labeled by two of the authors and discrepancies are discussed until a consensus. We provide the level of the inter-rater agreement in our qualitative analysis, and the values of the agreement are high (i.e., 0.72 and 0.81) in both qualitative studies.



\section{Related work}
\label{relatedwork}
\textbf{Leveraging user's feedback in Software Engineering}

\noindent{}To improve the quality of software systems, the software engineering community has proposed many approaches to leverage user's feedback. Since the feedback is based on first-hand experience from end-users, valuable insights can be mined to help developers identify patterns from issues and bugs. Khalid et al. studied user reviews from iOS apps and analyzed how user complaints, such as functional errors and app crashes, negatively affect app ratings in the Apple iOS App Store~\cite{Khalid:2015}. Hassan et al. analyzed negative app reviews due to app updates in the Google Play Store, and found that updates with feature removal and user interface issues lead to the highest increase of negative reviews ratio~\cite{Hassan:2018}. Mudambi et al. studied customer reviews on Amazon.com and identified the value of review extremity, review depth, and product type in helping customers make purchasing decisions~\cite{Mudambi:2010}. Poch{\'e} et al. found that around 30\% of user comments on YouTube videos for coding tutorials are useful to the original videos, and implemented an automated approach to identify useful comments for video creators~\cite{Poche:2017}. Lin et al. studied reviews on the Steam gaming platform, and found that players complain more about game design than software bugs~\cite{Lin:2018}.

Similar to previous studies that leverage user feedback in improving existing software artifacts, our study of comments under Stack Overflow answers investigates how comments add value to their associated answers. Furthermore, we examine whether informative comments are effectively presented using the current comment hiding mechanism.\\

\noindent\textbf{Improving the quality of knowledge sharing on Stack Overflow}

\noindent{}It is essential to deeply understand the knowledge sharing process on Stack Overflow, so that potential improvement can be made to benefit the Stack Overflow community. Wang et al. investigated how users interact with each other on Stack Overflow, and analyzed behaviors of both answer seekers and answerers to better understand how knowledge is formed to benefit individual users~\cite{Wang:2013}. Vasilescu et al. studied how users migrate questions from the R user support mailing list (r-help) to crowdsourced knowledge sharing platforms (i.e., Stack Overflow and Cross Validated)~\cite{Vasilescu:2014} and found that users can get faster answers on crowdsourced sites than on specialized mailing lists. Wang et al. analyzed how users revise answers on Stack Overflow under the current badge gamification system~\cite{Wang:2018} and provide some suggestions to improve their revising system. Ponzanelli et al. proposed an approach to identify low quality questions on Stack Overflow~\cite{Ponzanelli:2014a}. Srba et al. evaluated how low-quality content created by undesired groups of users on Stack Overflow negatively impacts the community, and proposed ways to solve the problems~\cite{Srba:2016}. In order to help users to find the right channel to ask questions, several approaches were developed to help users generate tags automatically when they post questions~\cite{WangLVS18,XiaLWZ13}. 

Our study also aims at improving the quality of content on Stack Overflow since informative comments under answers can add value to the knowledge sharing process. Different from prior studies that only focus on questions and answers on Stack Overflow, we focus on comments. We evaluate the efficacy of the comment hiding mechanism and provide actionable suggestions to help Stack Overflow improve the commenting system.\\

\noindent\textbf{Leveraging Comments on Stack Overflow}

\noindent\textbf{}Although prior studies of the Stack Overflow ecosystem mainly focus on questions and answers, some studies have taken comments into account. For example, in predicting the long-term value of question threads, Anderson et al. found that the number of comments in answers have significant predictive power~\cite{Anderson:2012}. 
Similarly, Tian et al. found that answers with more comments are more likely to be accepted~\cite{Tian:2013}. Asaduzzaman et al. analyzed both questions and their comments to find out why questions were unanswered~\cite{Asaduzzaman:2013}. They observed that users may post actual solutions in comments associated with these unanswered questions. Calefato et al. analyzed the sentiment of comments in their study of answer acceptance~\cite{Calefato:2015}. They found that the sentiment of comments significantly impacts the chance of answer acceptance. Dalip et al. observed that comment can provide additional information to improve the associated posts~\cite{Dalip:2013}. In addition, they found that commenting is useful for measuring the engagement of users in an answer, and this engagement improves the rating of answers. Chang et al. proposed a question routing framework to recommend answerers and commenters to a question~\cite{Chang:2013}. They observed the importance of commenting in further clarifications and the improvement of the quality of an answer. Zhang et al. leveraged comments to identify obsolete answers on Stack Overflow, and found that most observations of answer obsolescence in comments are supported with evidence~\cite{Zhang:2018}.

The above-mentioned studies extracted heuristic-based features from comments and observed the importance of commenting in the Stack Overflow ecosystem; however, no prior study has specifically investigated and characterized the phenomenon of commenting itself. In our current study, we study the informativeness of comments and the effectiveness of the current comment hiding mechanism. We wish to offer insights for users to better use Stack Overflow, and provide actionable suggestions for Stack Overflow engineers to enhance the current commenting system. 

\section{Conclusion}
\label{conclusion}
Stack Overflow uses a comment hiding mechanism where at most 5 comments are displayed under each answer. The goal of this mechanism is to improve the compactness of answer threads while retaining the informative comments to facilitate the knowledge sharing process. 40.5\% of comments are hidden by the comment hiding mechanism in these answers with hidden comments. 

In this study, we analyzed 1.3 million answers that have hidden comments to understand how the comment hiding mechanism classifies comments to display and hide. We found that more than half of hidden and displayed comments are informative. In addition, hidden comments are as informative as displayed comments, and these hidden comments even add a greater variety of informative content than displayed comments to their associated answers.

Furthermore, we evaluated the efficacy of the comment hiding mechanism and found that it fails to display informative comments. Comments are unfairly hidden due to the existence of tie-scored comments (especially 0-score comments). Finally, we provide a discussion on possible solutions to improve the comment hiding mechanism, such as replacing a longer unfairly hidden comment with a shorter unfairly displayed comment.

\bibliographystyle{IEEEtran}{

\begin{thebibliography}{10}
\providecommand{\url}[1]{#1}
\csname url@samestyle\endcsname
\providecommand{\newblock}{\relax}
\providecommand{\bibinfo}[2]{#2}
\providecommand{\BIBentrySTDinterwordspacing}{\spaceskip=0pt\relax}
\providecommand{\BIBentryALTinterwordstretchfactor}{4}
\providecommand{\BIBentryALTinterwordspacing}{\spaceskip=\fontdimen2\font plus
\BIBentryALTinterwordstretchfactor\fontdimen3\font minus
  \fontdimen4\font\relax}
\providecommand{\BIBforeignlanguage}[2]{{%
\expandafter\ifx\csname l@#1\endcsname\relax
\typeout{** WARNING: IEEEtran.bst: No hyphenation pattern has been}%
\typeout{** loaded for the language `#1'. Using the pattern for}%
\typeout{** the default language instead.}%
\else
\language=\csname l@#1\endcsname
\fi
#2}}
\providecommand{\BIBdecl}{\relax}
\BIBdecl

\bibitem{Zhang:2018}
\BIBentryALTinterwordspacing
H.~Zhang, S.~Wang, T.~P. Chen, Y.~Zou, and A.~E. Hassan, ``An empirical study
  of obsolete answers on {Stack Overflow},'' 2018. [Online]. Available:
  \url{https://doi.org/10.5281/zenodo.1320311}
\BIBentrySTDinterwordspacing

\bibitem{Gazan:2010}
R.~Gazan, ``Microcollaborations in a social {Q\&A} community,''
  \emph{Information Processing \& Management}, vol.~46, no.~6, pp. 693--702,
  2010.

\bibitem{Poche:2017}
E.~Poch{\'e}, N.~Jha, G.~Williams, J.~Staten, M.~Vesper, and A.~Mahmoud,
  ``Analyzing user comments on {YouTube} coding tutorial videos,'' in
  \emph{Proceedings of the 25th International Conference on Program
  Comprehension}, ser. ICPC '17, 2017, pp. 196--206.

\bibitem{Chang:2013}
S.~Chang and A.~Pal, ``Routing questions for collaborative answering in
  community question answering,'' in \emph{Proceedings of the 2013 IEEE/ACM
  International Conference on Advances in Social Networks Analysis and Mining},
  ser. ASONAM '13, 2013, pp. 494--501.

\bibitem{Kandogan:1998}
E.~Kandogan, ``Hierarchical multi-window management with elastic layout
  dynamics,'' Ph.D. dissertation, University of Michigan, 1998.

\bibitem{Cliff:1993}
N.~Cliff, ``Dominance statistics: Ordinal analyses to answer ordinal
  questions,'' \emph{Psychological bulletin}, vol. 114, no.~3, pp. 494--509,
  1993.

\bibitem{WangLXJ11}
S.~Wang, D.~Lo, Z.~Xing, and L.~Jiang, ``Concern localization using information
  retrieval: An empirical study on linux kernel,'' in \emph{18th Working
  Conference on Reverse Engineering, {WCRE} 2011, Limerick, Ireland, October
  17-20, 2011}, 2011, pp. 92--96.

\bibitem{WangLL14}
S.~Wang, D.~Lo, and J.~Lawall, ``Compositional vector space models for improved
  bug localization,'' in \emph{30th {IEEE} International Conference on Software
  Maintenance and Evolution, Victoria, BC, Canada, September 29 - October 3,
  2014}, 2014, pp. 171--180.

\bibitem{ThungWLL13}
F.~Thung, S.~Wang, D.~Lo, and J.~L. Lawall, ``Automatic recommendation of {API}
  methods from feature requests,'' in \emph{2013 28th {IEEE/ACM} International
  Conference on Automated Software Engineering, {ASE} 2013, Silicon Valley, CA,
  USA, November 11-15, 2013}, 2013, pp. 290--300.

\bibitem{Chen:2016}
T.-H. Chen, S.~W. Thomas, and A.~E. Hassan, ``A survey on the use of topic
  models when mining software repositories,'' \emph{Empirical Software
  Engineering}, pp. 1843--1919, 2016.

\bibitem{Oliveto:2010}
R.~Oliveto, M.~Gethers, D.~Poshyvanyk, and A.~D. Lucia, ``On the equivalence of
  information retrieval methods for automated traceability link recovery,'' in
  \emph{2010 IEEE 18th International Conference on Program Comprehension}, June
  2010, pp. 68--71.

\bibitem{Gethers:2011}
M.~Gethers, R.~Oliveto, D.~Poshyvanyk, and A.~D. Lucia, ``On integrating
  orthogonal information retrieval methods to improve traceability recovery,''
  in \emph{2011 27th IEEE International Conference on Software Maintenance},
  Sep. 2011, pp. 133--142.

\bibitem{Boslaugh:2012}
S.~Boslaugh, \emph{Statistics in a nutshell: a desktop quick reference}.\hskip
  1em plus 0.5em minus 0.4em\relax O'Reilly Media, 2012.

\bibitem{seaman1999qualitative}
C.~B. Seaman, ``Qualitative methods in empirical studies of software
  engineering,'' \emph{IEEE Transactions on Software Engineering (TSE)},
  vol.~25, no.~4, pp. 557--572, 1999.

\bibitem{seaman2008defect}
C.~B. Seaman, F.~Shull, M.~Regardie, D.~Elbert, R.~L. Feldmann, Y.~Guo, and
  S.~Godfrey, ``Defect categorization: making use of a decade of widely varying
  historical data,'' in \emph{Proceedings of the Second ACM-IEEE international
  symposium on Empirical software engineering and measurement}.\hskip 1em plus
  0.5em minus 0.4em\relax ACM, 2008, pp. 149--157.

\bibitem{Gwet:2002}
K.~Gwet, ``Inter-rater reliability: dependency on trait prevalence and marginal
  homogeneity,'' \emph{Statistical Methods for Inter-Rater Reliability
  Assessment Series}, vol.~2, no.~1, p.~9, 2002.

\bibitem{Xia2017}
X.~Xia, L.~Bao, D.~Lo, P.~S. Kochhar, A.~E. Hassan, and Z.~Xing, ``What do
  developers search for on the web?'' \emph{Empirical Software Engineering},
  pp. 3149--3185, 2017.

\bibitem{Calefato:2015}
F.~Calefato, F.~Lanubile, M.~C. Marasciulo, and N.~Novielli, ``Mining
  successful answers in {Stack Overflow},'' in \emph{Proceedings of the 12th
  Working Conference on Mining Software Repositories}, ser. MSR '15, 2015, pp.
  430--433.

\bibitem{mukaka2012guide}
M.~M. Mukaka, ``A guide to appropriate use of correlation coefficient in
  medical research,'' \emph{Malawi Medical Journal}, vol.~24, no.~3, pp.
  69--71, 2012.

\bibitem{Merton:1956}
R.~K. Merton, ``The matthew effect in science,'' \emph{Science}, vol. 159, no.
  3810, pp. 56--63, 1968.

\bibitem{Khalid:2015}
H.~Khalid, E.~Shihab, M.~Nagappan, and A.~E. Hassan, ``What do mobile app users
  complain about?'' \emph{IEEE Software}, vol.~32, no.~3, pp. 70--77, May 2015.

\bibitem{Hassan:2018}
S.~Hassan, C.~Bezemer, and A.~E. Hassan, ``Studying bad updates of top
  free-to-download apps in the google play store,'' \emph{IEEE Transactions on
  Software Engineering}, 2018.

\bibitem{Mudambi:2010}
S.~M. Mudambi and D.~Schuff, ``What makes a helpful online review? a study of
  customer reviews on {Amazon.com},'' \emph{MIS Quarterly}, vol.~34, no.~1, pp.
  185--200, 2010.

\bibitem{Lin:2018}
D.~Lin, C.-P. Bezemer, Y.~Zou, and A.~E. Hassan, ``An empirical study of game
  reviews on the steam platform,'' \emph{Empirical Software Engineering}, Jun
  2018.

\bibitem{Wang:2013}
S.~Wang, D.~Lo, and L.~Jiang, ``An empirical study on developer interactions in
  {StackOverflow},'' in \emph{Proceedings of the 28th Annual ACM Symposium on
  Applied Computing}, ser. SAC '13, 2013, pp. 1019--1024.

\bibitem{Vasilescu:2014}
B.~Vasilescu, A.~Serebrenik, P.~Devanbu, and V.~Filkov, ``How social {Q\&A}
  sites are changing knowledge sharing in open source software communities,''
  in \emph{Proceedings of the 17th ACM Conference on Computer Supported
  Cooperative Work \& Social Computing}, ser. CSCW '14, 2014, pp. 342--354.

\bibitem{Wang:2018}
S.~Wang, T.~P. Chen, and A.~E. Hassan, ``How do users revise answers on
  technical {Q\&A} websites? {A} case study on {Stack Overflow},'' \emph{IEEE
  Transactions on Software Engineering}, pp. 1--15, 2018.

\bibitem{Ponzanelli:2014a}
L.~Ponzanelli, A.~Mocci, A.~Bacchelli, M.~Lanza, and D.~Fullerton, ``Improving
  low quality {Stack Overflow} post detection,'' in \emph{2014 IEEE
  International Conference on Software Maintenance and Evolution}, 2014, pp.
  541--544.

\bibitem{Srba:2016}
I.~Srba and M.~Bielikova, ``Why is {Stack Overflow} failing? preserving
  sustainability in community question answering,'' \emph{IEEE Software},
  vol.~33, no.~4, pp. 80--89, July 2016.

\bibitem{WangLVS18}
S.~Wang, D.~Lo, B.~Vasilescu, and A.~Serebrenik, ``Entagrec ++: An enhanced tag
  recommendation system for software information sites,'' \emph{Empirical
  Software Engineering}, vol.~23, no.~2, pp. 800--832, 2018.

\bibitem{XiaLWZ13}
X.~Xia, D.~Lo, X.~Wang, and B.~Zhou, ``Tag recommendation in software
  information sites,'' in \emph{Proceedings of the 10th Working Conference on
  Mining Software Repositories, {MSR} '13, San Francisco, CA, USA, May 18-19,
  2013}, 2013, pp. 287--296.

\bibitem{Anderson:2012}
A.~Anderson, D.~Huttenlocher, J.~Kleinberg, and J.~Leskovec, ``Discovering
  value from community activity on focused question answering sites: A case
  study of {Stack Overflow},'' in \emph{Proceedings of the 18th ACM SIGKDD
  International Conference on Knowledge Discovery and Data Mining}, ser. KDD
  '12, 2012, pp. 850--858.

\bibitem{Tian:2013}
Q.~Tian, P.~Zhang, and B.~Li, ``Towards predicting the best answers in
  community-based question-answering services,'' in \emph{Proceedings of the
  7th International Conference on Weblogs and Social Media, {ICWSM} 2013,
  Cambridge, Massachusetts, USA, July 8-11, 2013.}, 2013.

\bibitem{Asaduzzaman:2013}
M.~Asaduzzaman, A.~S. Mashiyat, C.~K. Roy, and K.~A. Schneider, ``Answering
  questions about unanswered questions of {Stack Overflow},'' in
  \emph{Proceedings of the 10th Working Conference on Mining Software
  Repositories}, ser. MSR '13, 2013, pp. 97--100.

\bibitem{Dalip:2013}
D.~H. Dalip, M.~A. Gon\c{c}alves, M.~Cristo, and P.~Calado, ``Exploiting user
  feedback to learn to rank answers in q\&\#38;a forums: A case study with
  {Stack Overflow},'' in \emph{Proceedings of the 36th International ACM SIGIR
  Conference on Research and Development in Information Retrieval}, ser. SIGIR
  '13, 2013, pp. 543--552.

\end{thebibliography}
\footnotesize

}

\end{document}